%% file: article_R1.tex
\documentclass[prd,a4paper,preprint,preprintnumbers,nofootinbib,superscriptaddress]{revtex4-2}

\def\gGthree{\langle g^3 G^3 \rangle}
\def\gGfour{(\langle g^2 G^2 \rangle)^2}

\usepackage[a4paper, hdivide={1.91cm,,1.165cm}, vdivide={1.83cm,,3.0cm}]{geometry}

\input{my_preamble}

\usepackage{caption}
\allowdisplaybreaks[1]

\begin{document}


\title{Mixing angle between $^{3}P_1$ and $^{1}P_1$ states in heavy axial vector mesons within the QCD sum rules framework}

\author{T.~M.~Aliev\,\orcidlink{0000-0001-8400-7370}}
\email{taliev@metu.edu.tr}
\affiliation{Department of Physics, Middle East Technical University, Ankara, 06800, Turkey}

\author{S.~Bilmis\,\orcidlink{0000-0002-0830-8873}}
\email{sbilmis@metu.edu.tr}
\affiliation{Department of Physics, Middle East Technical University, Ankara, 06800, Turkey}
\affiliation{TUBITAK ULAKBIM, Ankara, 06510, Turkey}

\author{M.~Savci\,\orcidlink{0000-0002-6221-4595}}
\email{savci@metu.edu.tr}
\affiliation{Department of Physics, Middle East Technical University, Ankara, 06800, Turkey}

\date{\today}

\begin{abstract}

In this study, we calculate the mixing angles between the axial-vector mesons $D_{1(s1)} - D_{1(s1)}^\prime$ and $B_{1(s1)} - B_{1(s1)}^\prime$ using the QCD sum rules approach. Our results are $\theta_1 = 28.2 \pm 0.6^\circ$, $\theta_2 = 26.6 \pm 0.6^\circ$, $\theta_3 = 38.6 \pm 0.1^\circ$, and $\theta_4 = 38.5 \pm 0.1^\circ$. These values are in good agreement with the predictions of Heavy Quark Effective Theory, particularly for the mixing angle $\theta = 35.3^\circ$, and are compatible with several existing results in the literature. 

The predicted mixing angles can be tested through the analysis of semileptonic decays such as $B_c \to B_1 \ell \nu$, $B_c \to B_{s1}^0 \ell \nu$, $B_s \to D_{s1} \ell \nu$, and $B_s \to D_{s1}^\prime \ell \nu$, which can be investigated at experimental facilities such as LHCb and Belle II.

\end{abstract}

\maketitle


\newpage



\section{Introduction}
\label{intro}
 The quark model has been very successful in the classification of hadrons. According to the quark model, axial-vector mesons with quantum numbers $J^P=1^+$ are grouped into a nonet. In spectroscopic notation $n ^{2S+1} L_j$ there are two types of lowest p-wave mesons: $1^{3} P_1$ and $1^{1}P_1$ with C-parity $C=+1$ and $C=-1$, respectively. These states are usually denoted as $A_A$ and $A_B$. The physical mass eigenstates are mixtures of $A_A$ and $A_B$ states. The physical states $A$ and $A^\prime$ are defined in terms of $A_A$ and $A_B$ as follows:
 \begin{equation}
   \label{eq:1}
   \begin{split}
     A &= A_A \sin{\theta} + A_B \cos{\theta}  \\
     A^\prime &= A_A \cos{\theta} - A_B \sin{\theta}~.        
   \end{split}
 \end{equation}
 where $\theta$ is the mixing angle between $A_A$ and $A_B$ states.

 In this study, we focus our attention on charmed and bottom mesons with states $J^P = 1^+ $. Specifically, we consider the mixing between $D_1 - D_1^\prime (\theta_1)$, $D_{s1} - D_{s1}^\prime (\theta_2)$, $B_1 - B_1^\prime (\theta_3)$, and $B_{s1} - B_{s1}^\prime (\theta_4)$ which are listed with their associated masses in Table~\ref{tab:particles}. Note that although \(B_1\) and \(B_{s1}\) states have not been discovered yet, different theoretical approaches predicted slightly smaller mass splittings of $10 - 30~\rm{MeV}$ between $B_1$ and $B_1^\prime$ (see \cite{li:2021hss} and references therein).

\begin{table}[h!]
\centering
\begin{tabular}{ccc}
\toprule
\textbf{State-Pair}  & \textbf{States} & \textbf{Mass ($\rm{MeV}$)} \\
\midrule
\(D_{1} - D_{1}' ( \theta_1)\)    & \(D_1(2420)\)            & $2421.4 \pm 0.6$             \\
                     & \(D_1(2430)\)           & $2427  \pm 36$               \\
  \midrule
\(D_{s1} - D_{s1}' ( \theta_2) \) & \(D_{s1}(2460)\)    & $2459.5 \pm 0.6$                   \\
                     & \(D_{s1}(2536)\)  & $2535.1 \pm 0.1$                     \\
\midrule
  \(B_{1} - B_{1}' ( \theta_3)\) & \(B_1(5710)\)       &   $5710 $                   \\
                     & \(B_1(5721) \)             &    $5726.1 \pm 1.3$                   \\
  \midrule
\(B_{s1} - B_{s1}' ( \theta_4) \) & \(B_{s1} (5820)\)    &  $5820$                    \\
                     & \(B_{s1} (5830) \)       & $5828.7 \pm 0.4$                         \\
\bottomrule
\end{tabular}
\caption{The heavy-light mesons with $J^P = 1^+$~\cite{ParticleDataGroup:2022pth}. For the undiscovered states \(B_1'\) and \(B_{s1}'\), the mass values are adapted from theoretical predictions~\cite{li:2021hss}.}
\label{tab:particles}
\end{table}

The mixing angle is fundamental not only for understanding the nature of heavy axial-vector mesons but also for accurately determining their decay widths, which are critical for both theoretical studies and experimental verification.

One of the earliest comprehensive analyses of the mixing angles for $L=1$ mesons was conducted using the relativistic quark model~\cite{Godfrey:1986wj}. Since then, the mixing angle $\theta$ has been studied extensively using alternative approaches, including Bethe-Salpeter method~\cite{Godfrey:2015dva}, relativistic quark models~\cite{Godfrey:1986wj,Godfrey:2016nwn,Wang:2022tuy,Li:2018eqc,Sun:2014wea}, nonrelativistic quark models \cite{Lu:2016bbk, li:2021hss}, QCD potential models~\cite{Gupta:1994mw}, constituent quark model~\cite{Vijande:2004he,Yamada:2005nu}, chiral quark model~\cite{Zhong:2008kd}, Coulomb-gauge QCD model~\cite{Abreu:2019adi}, $^3P_0$ pair creation model~\cite{Ferretti:2015rsa}, and strong decay model~\cite{Close:2005se},  (see also review paper~\cite{Chen:2016spr}).

In this work, we calculate the mixing angles $\theta$ using the QCD sum rules approach, following the methodology first introduced in \cite{Aliev:2010ra}.
The paper is organized as follows. In Section~\ref{sec:2}, we derive the sum rules for the mixing angles of $1^{3}P_1$ and $1^{1}P_1$ states for heavy axial-vector mesons. In Section~\ref{sec:3}, we perform a numerical analysis of the sum rules for the mixing angles, and the final section contains our conclusion.
\section{Theoretical framework and derivation of mixing angles}
\label{sec:2}
In general, mass eigenstates do not coincide with flavor eigenstates. Hence, the mass eigenstates can be represented as linear combinations of flavor eigenstates. This implies that the interpolating currents for $A_i$ and \(A_i'\) mesons can be described as linear combinations of the currents for flavor eigenstates, i.e.,

\begin{equation}
  \label{eq:2}
  \begin{split}
    J_{A_i \mu} &= \sin{\theta_i} J_{A_i \mu}^{(0)} + \cos{\theta_i} J_{B_i \mu}^{(0)} \\
   J_{A'_i \mu} &= \cos{\theta_i} J_{A_i \mu}^{(0)} - \sin{\theta_i} J_{B_i \mu}^{(0)},
  \end{split}
\end{equation}
where the superscript \((0)\) denotes the current in the flavor eigenstates and
\begin{equation}
  \label{eq:4}
  \begin{split}
    J_{A_\mu}^{(0)} &= \bar{q} \gamma_\mu \gamma_5 Q, \quad \\
    J_{B_\nu}^{(0)} &= i \bar{q} \sigma_{\nu \alpha} p^\alpha \gamma_5 Q~,
  \end{split}
\end{equation}
with  \(q\) and \(Q\) representing the light and heavy quarks, respectively. Here  \(i = 1, 2, 3, 4\) correspond to the axial-vector \(D_{1}\),\(D_{s1}\), \(B_{1}\), and \(B_{s1}\) states. 

To determine the mixing angles \(\theta_i\), we follow the method presented in ~\cite{Aliev:2010ra} and start by considering the correlation function:

\begin{equation}
  \label{eq:3}
  \Pi_{\mu \nu}^{(A A')} (p) = \int d^4x \, e^{ipx} \langle 0 | T \{ J_{A_\mu}(x) \bar{J}_{A'_\nu}'(0) \} | 0 \rangle.
\end{equation}

To find the sum rules for the mixing angle \(\theta_i\), the correlation function is evaluated in two different domains. On one side, it is calculated in terms of hadrons by saturating a full set of hadrons carrying the same quantum numbers as the interpolating currents. On the other side, the correlation function is calculated in the deep Euclidean region (\(p^2 \ll 0\)) using the operator product expansion (OPE). These two representations are then matched, and to suppress higher states and the continuum, a Borel transformation with respect to the variable \(-p^2\) is performed. Since the currents \(J_{A \mu}\) and \(J_{A' \nu}'\) create only $A$ and $A'$ mesons from the vacuum, the hadronic part of the correlation function vanishes. Using Eqs. (\ref{eq:2}) and (\ref{eq:4}) from Eq.(\ref{eq:3} we get, 
\begin{equation}
  \label{eq:12}
  \cos{\theta} \sin{\theta} ( \Pi_{\mu \nu}^{(0)AA} - \Pi_{\mu \nu}^{0BB}) + (\cos^2{\theta}  - \sin^2{\theta}) \Pi_{\mu \nu}^{(0)AB} = 0~,
\end{equation}
where \(\Pi^{(0) i j}\) are the correlation functions for the unmixed states, i.e.,

\begin{equation}
\Pi_{\mu\nu}^{(0) i j} = i \int d^4x \, e^{ipx} \langle 0 | T \{ J_\mu^{(0)i}(x) \bar{J}_{\nu}^{(0)j} \} | 0 \rangle~.
\end{equation}

For axial-vector current, the correlation function can be expressed in terms of two independent invariant structures:

\begin{equation}
  \label{eq:6}
  \Pi_{\mu \nu}^{(0) i j }(p^2) = \Pi_1^{i j} (p^2) \left( g_{\mu \nu} - \frac{p_\mu p_\nu}{p^2} \right) + \Pi_2^{ i j} (p^2) \frac{p_\mu p_\nu}{p^2}.
\end{equation}

The structure \( g_{\mu \nu} - \frac{p_\mu p_\nu}{p^2} \) is associated with spin-1 particles. Thus,  we consider only this structure. Extracting  the coefficient of this structure from Eq. (\ref{eq:12}), the mixing angles are finally determined as: 
\begin{equation}
  \label{eq:5}
  \tan{2 \theta} = - \frac{2 \Pi_1^{ AB }}{\Pi_1^{AA} - \Pi_1^{BB}},
\end{equation}

It is worth noting that different conventions for the sign of mixing angles are used in the literature. This variation arises from the choice of quark ordering in representing the heavy-light quark system, which can be written either as \( \bar{q}Q \) or \( \bar{Q}q \). This difference in representation leads to opposite signs for the mixing angles. In our analysis, we adopt the convention presented in~\cite{Barnes:2002mu}, where the system is written as \( Q \bar{q} \).

With the conventions defined, we proceed to calculate the theoretical part of the correlation function in the deep Euclidean region (\( p^2 \ll 0 \)) using the OPE. The explicit expressions for the interpolating currents are employed, and Wick's theorem is applied.  After this operation, the correlation function is obtained in terms of quark propagators as follows:
\begin{equation}
  \label{eq:7}
  \begin{aligned}
    \Pi_{\mu \nu}^{(0)AA} (p^2) &= -i \int d^4x \, e^{i p x } \mathrm{Tr} \left[ S_q^{ab}(-x) \gamma_\mu \gamma_5 S_Q^{ba}(x) \gamma_\nu \gamma_5 \right], \\
    \Pi_{\mu \nu}^{(0)AB} (p^2) &= -i \int d^4x \, e^{i p x } \mathrm{Tr} \left[ S_q^{ab}(-x) \sigma_{\mu \alpha} p^\alpha \gamma_5 S_Q^{ba}(x) \gamma_\nu \gamma_5 \right], \\
    \Pi_{\mu \nu}^{(0)BB} (p^2) &= -i \int d^4x \, e^{i p x } \mathrm{Tr} \left[ S_q^{ab}(-x) \sigma_{\mu \alpha} p^\alpha \gamma_5 S_Q^{ba}(x) \sigma_{\nu \beta} p^\beta \gamma_5 \right].
  \end{aligned}
\end{equation}
The quark propagators for the light and heavy quarks in the  \(x\) representation are given by Eqs.~\eqref{eq:7} and \eqref{eq:8}, respectively, where \(G^{\mu\nu}\) is the gluon field strength tensor, and \(K_i\) are the modified Bessel functions of the second kind (see \cite{Huang:2012ti, ParticleDataGroup:2022pth}).

\begin{equation}
  \label{eq:8}
\begin{aligned}
iS_q^{ab}(x) &= \frac{i \delta^{ab}}{2\pi^2 x^4} \not\!x
   - \frac{\delta^{ab}}{12} \langle \bar{q}q \rangle
   - \frac{i \delta^{ab} g_s^2 x^2 \not\!x}{2^5 3^5} \langle \bar{q}q \rangle^2 \\
&\quad + \frac{i}{32\pi^2} g_s G_{\mu\nu}^{ab}
   \frac{\sigma^{\mu\nu}\not\!x + \not\!x \sigma^{\mu\nu}}{x^2}
   + \frac{\delta^{ab} x^2}{192} \langle g_s \bar{q}\sigma G q \rangle \\
&\quad - \frac{\delta^{ab} x^4}{2^{10} 3^3} \langle \bar{q}q \rangle \langle g_s^2 G^2 \rangle
   + \cdots \\
&\quad - \frac{m_q \delta^{ab}}{4\pi^2 x^2}
   + \frac{i m_q \delta^{ab} \not\!x}{48} \langle \bar{q}q \rangle
   - \frac{m_q \delta^{ab} g_s^2 x^4}{2^7 3^5} \langle \bar{q}q \rangle^2 \\
&\quad + \frac{m_q}{32\pi^2} g_s G_{\mu\nu}^{ab} \sigma^{\mu\nu} \ln(-x^2)
   - \frac{i m_q \delta^{ab} x^2 \not\!x}{2^7 3^2} \langle g_s \bar{q}\sigma G q \rangle \\
&\quad - \frac{m_q \delta^{ab}}{2^9 3\pi^2} x^2 \ln(-x^2) \langle g_s^2 G^2 \rangle
   + \cdots,
\end{aligned}
\end{equation}
%
%
\begin{equation}
  \label{eq:9}
\begin{aligned}
S_Q^{ab}(x) &= \frac{m_Q^2 \delta^{ab}}{(2 \pi)^2} \left[
i \rlap{/}x \frac{K_2(m_Q \sqrt{-x^2})}{(\sqrt{-x^2})^2} +
\frac{K_1(m_Q \sqrt{-x^2})}{\sqrt{-x^2}} \right] \\
&\quad - \frac{m_Q g_s G_{\mu\nu}^{ab}}{8 (2\pi)^2} \left[
i (\sigma^{\mu\nu} \rlap{/}x + \rlap{/}x \sigma^{\mu\nu}) 
\frac{K_1(m_Q \sqrt{-x^2})}{\sqrt{-x^2}} +
2 \sigma^{\mu\nu} K_0(m_Q \sqrt{-x^2}) \right] \\
&\quad - \frac{\langle g_s^2 G^2 \rangle \delta^{ab}}{2^6 3^2 (2 \pi)^2} \left[
(i m_Q \rlap{/}x - 6) \frac{K_1(m_Q \sqrt{-x^2})}{\sqrt{-x^2}} +
m_Q x^4 \frac{K_2(m_Q \sqrt{-x^2})}{(\sqrt{-x^2})^2} \right] \\
&\quad + \frac{\langle g_s^3 G^3 \rangle \delta^{ab}}{2^8 3^2 (2 \pi)^2} \left[
- \frac{i \rlap{/}x x^2}{m_Q} \frac{K_1(m_Q \sqrt{-x^2})}{\sqrt{-x^2}} +
i \rlap{/}x x^4 \frac{K_2(m_Q \sqrt{-x^2})}{(\sqrt{-x^2})^2} \right. \\
&\quad + \left. \frac{10}{m_Q} x^4 \frac{K_2(m_Q \sqrt{-x^2})}{(\sqrt{-x^2})^2} +
x^4 \frac{K_1(m_Q \sqrt{-x^2})}{\sqrt{-x^2}} \right]~.
\end{aligned}
\end{equation}
%

The invariant function for the Lorentz structure $g_{\mu \nu} - \ds{ p_\mu p_\nu \over p^2}$ can be expressed with its imaginary part (spectral density) via the dispersion relation:
\begin{equation}
  \label{eq:10}
  \Pi_1^{ i j}(p^2) = \int_{m_Q^2}^\infty ds \, \frac{\rho_1^{ij}(s)}{s - p^2}.
\end{equation}
Using the explicit expressions for the light and heavy quark propagators, the spectral density can be calculated straightforwardly.  The expressions for the spectral densities are presented in the Appendix. After performing Borel transformation over the variable \((-p^2)\) and imposing quark-hadron duality, we get

\begin{equation}
  \label{eq:11}
  \Pi_1^{ij}(M^2) = \int_{m_Q^2}^{s_0} ds \, \rho_1^{ij}(s) e^{-s /M^2},
\end{equation}
where \(s_0\) is the continuum threshold in the corresponding channel.
Finally, substituting these results into Eq.\eqref{eq:5}, we can determine the mixing angle.
\section{Numerical Analysis}
\label{sec:3}

After having established the theoretical framework, we conduct a numerical analysis of the sum rules to determine the mixing angle in this section. We begin by listing the input parameters used in the sum rules in Table \ref{tab:1}. The heavy quark masses are given in the $\overline{\text{MS}}$ scheme, while the values of strange quark mass and the quark condensate are presented at $\mu = 1~\rm{GeV}$ scale. 

\begin{table}[hbt]
\centering
\begin{adjustbox}{center}
\renewcommand{\arraystretch}{1.2}
\setlength{\tabcolsep}{6pt}
\begin{tabular}{lr}
\toprule
$\overline{m}_s (1~\rm{GeV})$      & $0.126~\rm{GeV}$~\cite{ParticleDataGroup:2022pth} \\
$\overline{m}_b (\overline{m}_b)$  & $4.18^{+0.03}_{-0.02}~\rm{GeV}$~\cite{ParticleDataGroup:2022pth} \\
$\overline{m}_c (\overline{m}_c)$  & $(1.27 \pm 0.02)~\rm{GeV}$~\cite{ParticleDataGroup:2022pth} \\
$\langle \bar{q}q \rangle~\rm{(1~GeV)}$ & $(-1.65 \pm 0.15)\times 10^{-2}~\rm{GeV}^3$~\cite{Ioffe:2005ym} \\
$\langle \bar{s}s \rangle$         & $(0.8 \pm 0.2) \langle \bar{q}q \rangle~\rm{GeV^3}$~\cite{Ioffe:2005ym} \\
$m_0^2$                            & $(0.8 \pm 0.2)~\rm{GeV^2}$~\cite{Ioffe:2005ym} \\
$\langle g^2 G^2 \rangle$ & $ 4 \pi^2 (0.012 \pm 0.006)~\rm{GeV^4}$~\cite{Ioffe:2002ee} \\
$\langle g^3 G^3 \rangle$ & $(0.57 \pm 0.29)~\rm{GeV^6}$~\cite{Narison:2022paf} \\
  \bottomrule
\end{tabular}
\end{adjustbox}
\caption{The numerical values of the input parameters.}
\label{tab:1}
\end{table}

\begin{table}[hbt]
\centering
\begin{adjustbox}{center}
\renewcommand{\arraystretch}{1.2}
\setlength{\tabcolsep}{6pt}
\begin{tabular}{lcc}
\toprule
                & \(M^2\)  (\(\rm{GeV}^2\)) & \(s_0\) (\(\rm{GeV}^2\))  \\
\midrule
\(D_1\)         & \(3 \le M^2 \le 6\)    & \(9 \pm 1\)   \\
\(D_{s1}\)    & \(3 \le M^2 \le 6\)    &  \(10 \pm 1\)  \\
\(B_1\)         & \(8 \le M^2 \le 12\)   & \(41 \pm 1\)  \\
\(B_{s1}\)    & \(9 \le M^2 \le 13\)   & \(43 \pm 1\)  \\
\bottomrule
\end{tabular}
\end{adjustbox}
\caption{The working regions for the Borel mass parameter \(M^2\) and the continuum threshold \(s_0\).}
\label{tab:region}
\end{table}

\begin{table}[hbt]
\centering
\begin{adjustbox}{center}
\renewcommand{\arraystretch}{1.2}
\setlength{\tabcolsep}{6pt}
\begin{tabular}{lcccc}
\toprule
Reference & $\theta_1$ & $\theta_2$ & $\theta_3$ & $\theta_4$ \\
\midrule
Present Work            & $28.2 \pm 0.6^\circ$ & $26.6 \pm 0.6^\circ$ & $38.6 \pm 0.1^\circ$ & $38.5 \pm 0.1^\circ$ \\
\cite{li:2021hss}       & ---                  & ---                  & $-35.2^\circ$        & $-39.6^\circ$        \\
\cite{Godfrey:2015dva}  & $-25.7^\circ$        & $-37.5^\circ$        & ---                  & ---                  \\
\cite{Wang:2022tuy}     & $-58.3 \pm 9.0^\circ$& ---                  & --- & ---                  \\
\cite{Li:2018eqc}       & $35.1^\circ$         & $-60.4^\circ$        & $-55.4^\circ$        & $-55.3^\circ$        \\
\cite{Lu:2016bbk}       & ---                  & ---                  & $-34.6^\circ$        & $-34.9^\circ$        \\
\cite{Gupta:1994mw}     & $29.0^\circ$         & $26.0^\circ$         & $31.7^\circ$         & $27.3^\circ$         \\
\cite{Vijande:2004he}   & $43.5^\circ$         & $58.4^\circ$         & ---                  & ---                  \\
\cite{Yamada:2005nu}    & ---                  & $-45.4^\circ$        & ---                  & ---                  \\
\cite{Abreu:2019adi}    & $34.0^\circ$         & $33.0^\circ$         & $35.0^\circ$         & $34.8^\circ$         \\
\cite{Ferretti:2015rsa} & $25.7^\circ$         & $37.5^\circ$         & $30.3^\circ$         & $39.1^\circ$         \\
\cite{Close:2005se}     & $-54.7^\circ (35.3^\circ)$ & $-54.7^\circ (35.3^\circ)$ & $-54.7^\circ (35.3^\circ)$ & $-54.7^\circ (35.3^\circ)$ \\
\cite{Godfrey:1986wj}   & $-26.0^\circ$        & $-38.0^\circ$        & $-31.0^\circ$        & $-40.0^\circ$        \\
\cite{Godfrey:2016nwn}  & ---                  & ---                  & $30.3^\circ (43.6^\circ)$ & $39.1^\circ (37.9^\circ)$ \\
\cite{Sun:2014wea}      & ---                  & ---                  & $-73.5 \pm 3.5^\circ \; (-36.5 \pm 3.5^\circ)$ & --- \\
\cite{Zhong:2008kd}     & $-55^\circ$          & $-55^\circ$          & ---                  & ---                  \\
\bottomrule
\end{tabular}
\end{adjustbox}
\caption{Comparison of mixing angles between the heavy axial-vector mesons. Here $\theta_1$, $\theta_2$, $\theta_3$, and $\theta_4$ describe the mixing angles between the $D_{1(s1)} - D_{1(s1)}^\prime$ and $B_{1(s1)} - B_{1(s1)}^\prime$ states, respectively.}
\label{tab:combined_table}
\end{table}

In addition to these input parameters, there are two  auxiliary parameters in the sum rules: the Borel mass parameter \(M^2\) and the continuum threshold \(s_0\). Table~\ref{tab:region} summarizes the working regions of these parameters, which are chosen to ensure the stability and convergence of the sum rules.

Since \(M^2\) and \(s_0\) are auxiliary parameters, the mixing angle should be independent of them. The working region of \(M^2\), where the mixing angle is very weakly dependent on these parameters, is determined by imposing specific conditions.

The upper limit of \(M^2\) is determined by demanding that higher state and continuum contributions remain below 40\% of the total result. This condition can be expressed as:

\begin{equation}
{\int_{m_Q^2}^{s_0} ds \, \rho_1(s) e^{-s/M^2} \over \int_{m_Q^2}^{\infty} ds \, \rho_1(s) e^{-s/M^2} } \geq 0.6~.
\end{equation}

The lower limit of \(M^2\) is determined by requiring that the operator product expansion (OPE) should be convergent. Specifically, the contribution of the highest dimensional condensate must be less than 10\% of the total result. The continuum threshold \(s_0\) is selected to minimize the variation of the mixing angle within the Borel mass working region. These conditions lead to the following working regions for \(M^2\) and \(s_0\), as summarized in Table \ref{tab:region}.


\begin{figure}
	\centering
	\begin{subfigure}{0.4\textwidth} 
		\includegraphics[width=\textwidth]{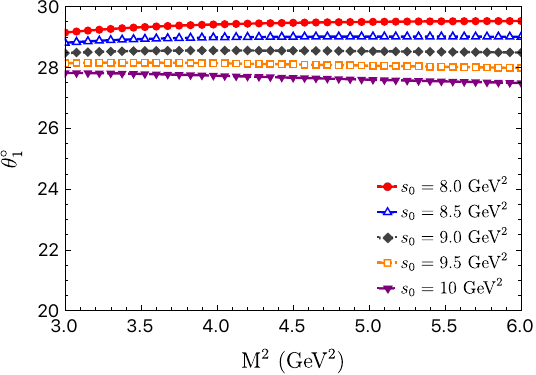}
	\end{subfigure}
	\vspace{1em} 
	\begin{subfigure}{0.4\textwidth} 
		\includegraphics[width=\textwidth]{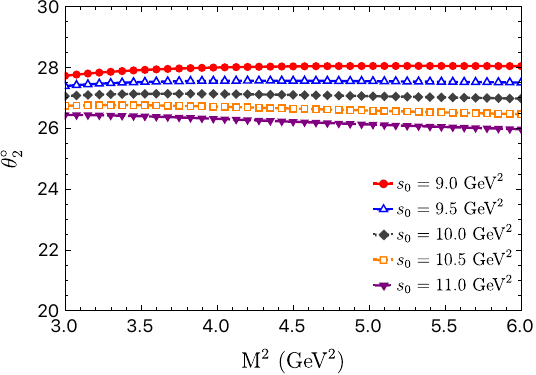}
	\end{subfigure}
	\begin{subfigure}{0.4\textwidth} 
		\includegraphics[width=\textwidth]{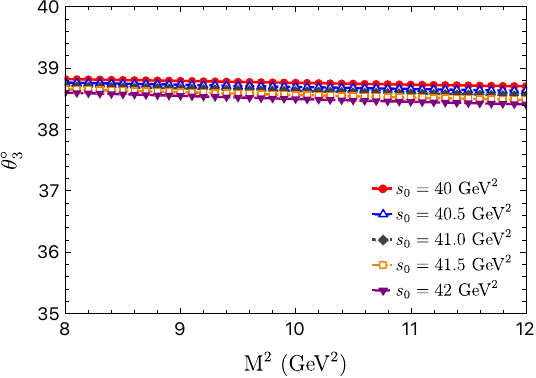}
	\end{subfigure}
	\begin{subfigure}{0.4\textwidth} 
		\includegraphics[width=\textwidth]{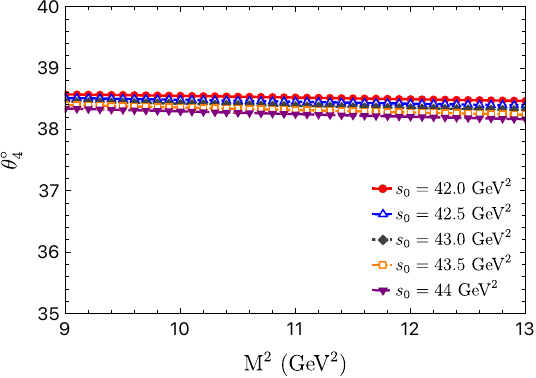}
	\end{subfigure}
	\caption{Dependency of mixing angles on $M^2$ at several fixed values $s_0$: (a)$D_1 - D_1^\prime$, (b)$D_{s1} - D_{s1}^\prime$, (c) $B_1 - B_1^\prime$, (d) $B_{s1} - B_{s1}^\prime$. } 
        \label{fig:1}
\end{figure}


\begin{figure}
	\centering
	\begin{subfigure}{0.4\textwidth} 
		\includegraphics[width=\textwidth]{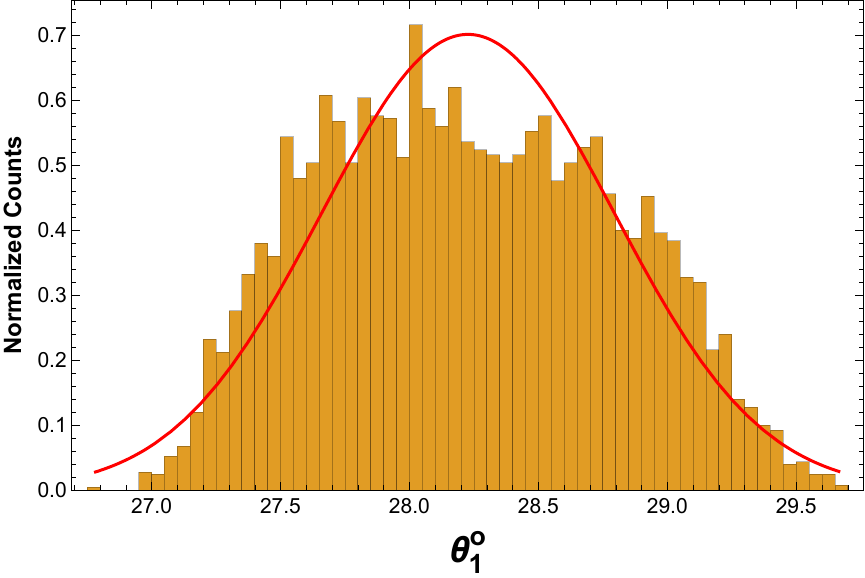}
	\end{subfigure}
	\vspace{1em} 
	\begin{subfigure}{0.4\textwidth} 
		\includegraphics[width=\textwidth]{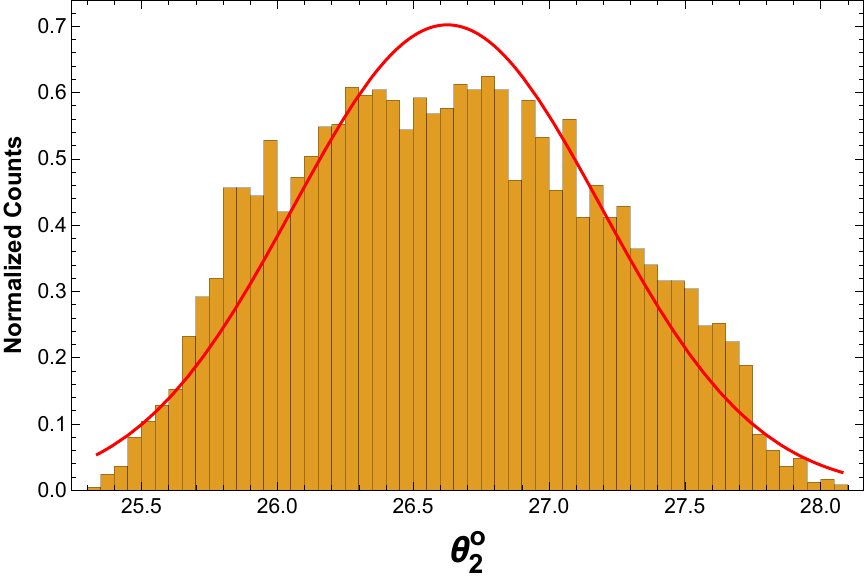}
	\end{subfigure}
	\begin{subfigure}{0.4\textwidth} 
		\includegraphics[width=\textwidth]{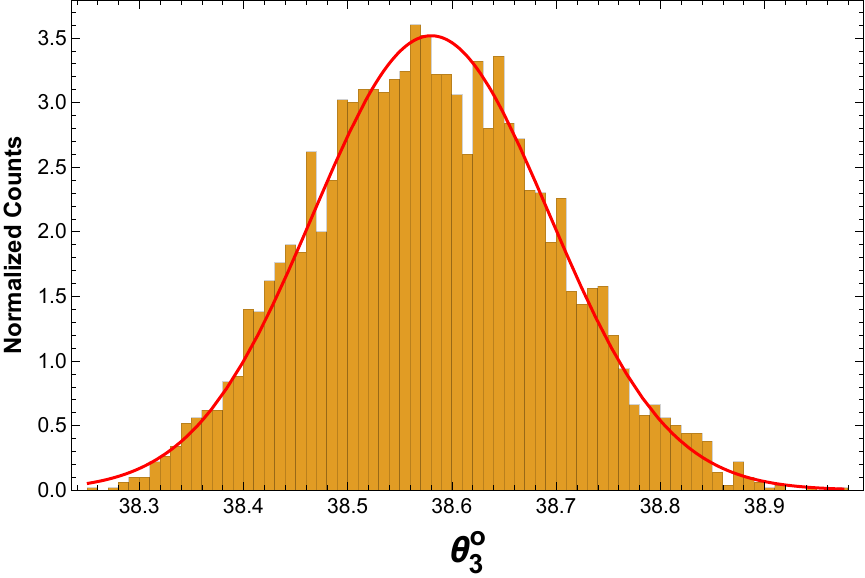}
	\end{subfigure}
	\begin{subfigure}{0.4\textwidth} 
		\includegraphics[width=\textwidth]{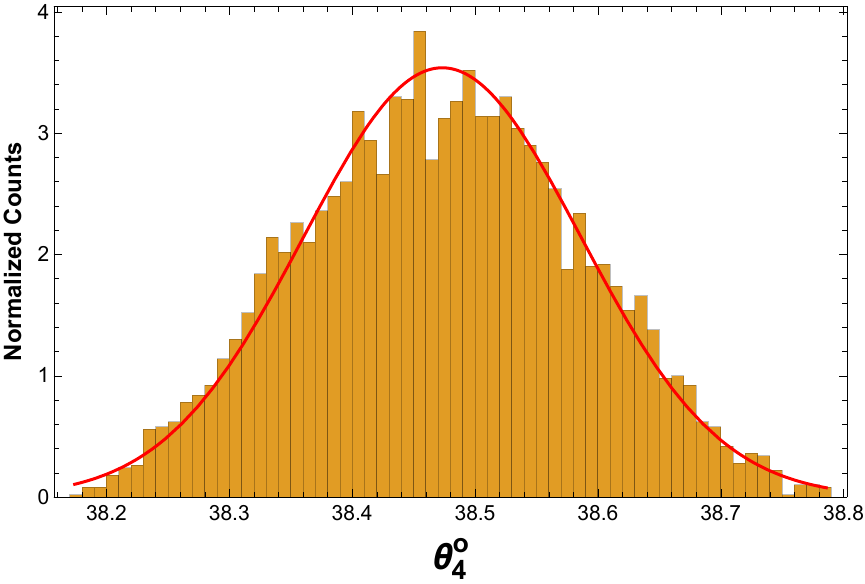}
	\end{subfigure}
	\caption{Distribution of normalized counts for the mixing angles \(\theta_1\), \(\theta_2\), \(\theta_3\), and \(\theta_4\), obtained through a Monte Carlo analysis to determine their uncertainties. The input parameters were randomly varied within their uncertainties. The histograms represent the normalized counts, and the red curves show Gaussian fits to the resulting distributions: (a)$D_1 - D_1^\prime$, (b)$D_{s1} - D_{s1}^\prime$, (c) $B_1 - B_1^\prime$, (d) $B_{s1} - B_{s1}^\prime$.}
        \label{fig:2}
\end{figure}

Having determined the working regions of $M^2$ and $s_0$, the mixing angles $\theta$ between the axial-vector mesons $D_{1(s1)} - D_{1(s1)}^\prime$ and $B_{1(s1)} - B_{1(s1)}^\prime$  are evaluated. Their dependence on $M^2$, at several fixed $s_0$ values are presented in Fig. \ref{fig:1}.

To account for uncertainties in the input parameters, as well as the auxiliary parameters $M^2$ and $s_0$, we performed a Monte Carlo analysis. By randomly sampling these parameters  5000 times within their respective uncertainty ranges, we generated statistical distributions for the mixing angles. The central values and uncertainties were extracted through Gaussian fits to these distributions (see Fig.~\ref{fig:2}). 

The values of the mixing angles obtained from Monte Carlo analysis are collected in Table \ref{tab:combined_table}. For completeness, we present the predictions of other approaches existing in the literature.

Our results for $\theta_1 = 28.2 \pm 0.6^\circ$, $\theta_2 = 26.6 \pm 0.6^\circ$, $\theta_3 = 38.6 \pm 0.1^\circ$, and $\theta_4 = 38.5 \pm 0.1^\circ$ are notably close to the positive angle predicted by heavy quark effective theory $\theta = +35.3^\circ$~\cite{Barnes:2002mu}. The small deviations can be attributed to finite heavy quark mass corrections.

A key advantage of the method used in this work is that the mixing angles are determined only by QCD parameters and are free from hadronic degrees of freedom. This independence is particularly important, as many alternative approaches rely on meson mass inputs that are not yet well determined experimentally, introducing  uncertainties into their predictions. 
\section{Conclusion}
\label{sec:4}
In this study, we calculated the mixing angles between the axial-vector mesons $D_{1(s1)} - D_{1(s1)}^\prime$ and $B_{1(s1)} - B_{1(s1)}^\prime$ within the framework of QCD sum rules. Our results are $\theta_1 = 28.2 \pm 0.6^\circ$, $\theta_2 = 26.6 \pm 0.6^\circ$, $\theta_3 = 38.6 \pm 0.1^\circ$, and $\theta_4 = 38.5 \pm 0.1^\circ$, and are in good agreement with the predictions of heavy quark effective theory in the heavy quark limit, for the mixing angle $\theta = 35.3^\circ$. Furthermore, our results show compatibility with several existing studies in the literature. A key advantage of our approach  is its independence from hadronic degrees of freedom.

Our predictions for the  mixing angles can be tested through the analysis of semileptonic decays of $B$ and $B_c$ mesons. Specifically, decays such as $B_c \to B_1 \ell \nu$, $B_c \to B_{s1}^0 \ell \nu$, $B_s \to D_{s1} \ell \nu$, and $B_s \to D_{s1}^\prime \ell \nu$ offer promising areas for experimental verification at facilities such as LHCb and Belle II. 

The discrepancies observed between our results and those obtained using other theoretical approaches highlight the need for further theoretical refinements and experimental investigations.



\bibliographystyle{utcaps_mod}
\bibliography{all.bib}


\newpage

\appendix*
\section{The expression of the spectral densities}
\label{app:formulas}
Spectral densities corresponding to {\boldmath$g_{\mu\nu}$} structure in the polarization operator.
\begin{equation}
\begin{aligned}
\rho_1^{AA} - \rho_1^{BB} &= e^{-m_Q^2/M^2} \Bigg\{ \\
&\quad  \frac{1}{2^9 3^4 M^{14}} m_Q^5 \Big[  
 4 \gGfour m_Q^2 + 24 m_0^2 m_q m_Q \gGthree \Big] \qq \\
&\quad - \frac{1}{2^9 3^4 M^{12}} m_Q^3 \Big[ 
\Big( 52 \gGfour - 72 \gGthree m_0^2 \Big) m_Q^2 \\
&\quad + 48 m_0^2 m_q m_Q \Big( 4 \gGthree + \gGgG m_Q^2 \Big) \Big] \qq \\
&\quad + \frac{1}{2^8 3^3 M^{10}} m_Q \Big\{ 
2 m_Q \Big[ m_Q \Big( 14 \gGfour - 51 \gGthree m_0^2 -
12 \gGgG m_0^2 m_Q^2 \Big) \\
&\quad + 6 m_q \gGthree (m_0^2 + 2 m_Q^2)  
- 48 \gGgG m_0^2 m_q m_Q^2 \Big] \Big\} \qq \\  
&\quad - \frac{1}{2^6 3^3 M^8}
\Big\{ 5 \gGgG - 6 m_Q \Big[ 7 m_0^2 m_Q - 2m_Q^3 -
2 m_q (6 m_0^2 - m_Q^2) \Big] \Big\} m_Q \gGgG \qq \\
&\quad + \frac{1}{2^8 3^3 m_Q M^8} \Big\{ 
6 m_Q \gGthree \Big[ 23 m_0^2 m_Q - 8 m_Q^3 + m_q (30 m_0^2 + 8 m_Q^2) \Big] \Big\} \qq \\ 
&\quad - \frac{1}{144 M^6}
\Big\{ \gGgG \Big[ 4 m_0^2 m_Q - 6m_Q^3 -
m_q (4 m_0^2 - 7 m_Q^2) \Big] - 12 m_0^2 m_q m_Q^4 \Big\} \qq \\
&\quad + \frac{1}{288 m_Q M^6} \gGthree \Big[ 2 m_0^2 + 12 m_q m_Q + 7 m_Q^2 \Big] \qq \\
&\quad - \frac{1}{48 M^4} 
\Big[ 2m_Q \Big(\gGgG - 6 m_0^2 m_Q^2\Big) - 3 m_q \Big(\gGgG - 4 m_0^2
m_Q^2\Big) \Big] \qq \\
&\quad + \frac{1}{192 m_Q M^4 \pi^2} \gGthree \Big(m_q m_Q^2 + 2 \pi^2 \qq\Big) \\
&\quad - \frac{1}{96 M^2 \pi^2}
m_Q \Big[ \gGgG m_q + 24 \pi^2 (m_0^2 + 2 m_q m_Q) \qq \Big] \\
&\quad - \frac{1}{384 m_Q M^2 \pi^2} \gGthree (m_q + 2 m_Q) \\
&\quad + \frac{1}{96 m_Q \pi^2}
\Big[ \gGgG (m_q + m_Q) - 24 \pi^2 \qq (m_0^2 - 2 m_q m_Q + 4 m_Q^2) \Big] \\
&\quad - \frac{1}{384 m_Q^3 \pi^2} \gGthree (3 m_q + 8 m_Q)
\Bigg\} \\
&\quad - \int_{m_Q^2}^{s_0} ds\,
e^{-s/M^2} \Bigg\{
\frac{1}{32 s \pi^2} \gGgG +
\frac{3}{8 m_Q s \pi^2} \Big\{ (s - m_Q^2) \Big[ s m_Q - m_Q^3 - m_q (s + m_Q^2) \Big] \Big\} 
\Bigg\}~.
\end{aligned}
\end{equation}

\begin{equation}
\begin{aligned}
\rho_1^{AB} &= e^{-m_Q^2/M^2} \Bigg\{ \\
&\quad - \frac{1}{2^8 3^4 M^{12}} m_Q^3 \Big[
 14 \gGfour m_Q^2 + 96 \gGthree m_0^2 m_q m_Q \Big] \qq \\
&\quad - \frac{1}{2^7 3^3 M^{10}} m_Q \Big[
 - \Big( 22 \gGfour - 36 \gGthree m_0^2 \Big) m_Q^2 \\
&\quad - 7 m_0^2 m_q m_Q \Big(17 \gGthree + 4 \gGgG m_Q^2 \Big) \Big] \qq \\
&\quad - \frac{1}{2^6 3^3 M^8} \gGgG m_Q \Big\{ 25 \gGgG - 6 m_Q \Big[ 5 m_0^2 m_Q -
m_q (14 m_0^2 - 4 m_Q^2) \Big] \Big\} \qq \\
&\quad + \frac{1}{2^6 3^2 m_Q M^8} \gGthree \Big[ 5 \Big(\gGgG + m_0^2 m_Q^2 \Big) -
m_q m_Q (23 m_0^2 - 10 m_Q^2) \Big] \qq \\
&\quad + \frac{1}{432 m_Q M^6}
\Big[ 3 \gGgG \Big(\gGgG - 6 m_0^2 m_Q^2\Big) +
m_q m_Q \gGgG (20 m_0^2 - 39 m_Q^2) + 72 m_0^2 m_q m_Q^5 \Big] \qq \\
&\quad + \frac{1}{2^6 3^3 m_Q^2 M^6} \gGthree \Big[ 108 m_0^2 m_Q - m_q (20 m_0^2 + 81 m_Q^2) \Big] \qq \\
&\quad - \frac{1}{96 m_Q^2 M^4} m_q \Big[ 3 \gGthree - 
4 m_Q^2 \Big(3 \gGgG - 14 m_0^2 m_Q^2 \Big) \Big] \qq \\
&\quad + \frac{1}{12 m_Q M^2}
\Big\{ 2 \gGgG - 3 m_Q \Big[ 4 m_0^2 m_Q - m_q (m_0^2 + 4 m_Q^2) \Big] \Big\} \qq \\
&\quad - \frac{1}{96 m_Q M^2 \pi^2} \gGthree (2 m_q - m_Q) \\
&\quad - \frac{1}{96 m_Q^2 \pi^2}
\Big\{ 35 \gGgG m_Q^2 - 
2 m_q \Big[ \gGgG m_Q + 8 \pi^2 (m_0^2 + 9 m_Q^2) \qq \Big] \Big\} \\
&\quad - \frac{1}{96 m_Q^3 \pi^2} \Big[ \gGthree (5 m_q - 2 m_Q) \Big]   
\Bigg\} \\
&\quad + \int_{m_Q^2}^{s_0} ds\,
e^{-s/M^2} \Bigg\{
\frac{1}{48 s^2 \pi^2} \gGgG (4 s - m_Q^2) -
\frac{1}{384 m_Q^2 s^2 \pi^2} \gGthree (2 s + m_Q^2) \\
&\quad - \frac{1}{8 \pi^2 m_Q^2 s^2}
(s - m_Q^2)^2 (s^2 - 6 s m_q m_Q - m_Q^4)
\Bigg\}~.
\end{aligned}
\end{equation}




\end{document}

%% file: my_preamble.tex
\usepackage[hyperfootnotes=false]{hyperref}
\hypersetup{linkcolor=red,
citecolor=green,
filecolor=cyan,
urlcolor=magenta}

\usepackage{amsfonts}

\usepackage{amsmath}
\usepackage{amssymb}

\usepackage{bbm}

\usepackage{bbold}

\usepackage{adjustbox}
\usepackage{booktabs}

\usepackage{colortbl}

\usepackage{cancel}

\usepackage{color}
\usepackage{xcolor}

\usepackage{graphicx}
\usepackage{xspace}

\usepackage{units}

\usepackage{slashed}

\usepackage{tensor}

\usepackage{gensymb}

\usepackage{cleveref}

\usepackage{tabularx}
\usepackage{multirow}

\usepackage{setspace}
\usepackage{orcidlink}

\usepackage{caption}
\usepackage{subcaption}

\usepackage{enumerate}
\usepackage{enumitem}

\usepackage[version=3]{mhchem}

\usepackage{lipsum}

\newcommand{{\ia} }{{\i}}
\newcommand{{\Ia} }{{\.I}}

\newcommand{\ra}{\rightarrow}
\newcommand{\la}{\leftarrow}

\def\s2tw{{\rm sin ^2 \theta_{W}}}

\def\ds{\displaystyle}
\def\beq{\begin{equation}}
\def\eeq{\end{equation}}
\def\bea{\begin{eqnarray}}
\def\eea{\end{eqnarray}}

\def\la{\langle}
\def\ra{\rangle}

\def\qq{\la \bar q q \ra}
\def\gGgG{\la g^2 G^2 \ra}

\def\qq{\langle \bar q q \rangle}

\def\ds{\displaystyle}
\def\bea{\begin{eqnarray}}
\def\eea{\end{eqnarray}}
\def\beeq{\begin{eqnarray}}
\def\eeeq{\end{eqnarray}}

\def\la{\langle}
\def\ra{\rangle}
\def\ba{\begin{array}}
\def\ea{\end{array}}

\def\xis0{{\Xi^{*0}}}

\def\qq{\la \bar q q \ra}
\def\gGgG{\la g^2 G^2 \ra}

\def\g5{\gamma_5}




%% file: article_R1.bbl
\providecommand{\href}[2]{#2}\begingroup\raggedright\begin{thebibliography}{10}

\bibitem{li:2021hss}
Q.~li, R.-H. Ni, and X.-H. Zhong, ``{\em {Towards establishing an abundant $B$
  and $B_s$ spectrum up to the second orbital excitations}},''
  \href{http://dx.doi.org/10.1103/PhysRevD.103.116010}{Phys. Rev. D {\bfseries
  103} (2021) 116010}, \href{http://arxiv.org/abs/2102.03694}{[{\ttfamily
  2102.03694}]}.

\bibitem{ParticleDataGroup:2022pth}
{\bfseries Particle Data Group} Collaboration, R.~L. Workman {\em et~al.},
  ``{\em {Review of Particle Physics}},''
  \href{http://dx.doi.org/10.1093/ptep/ptac097}{PTEP {\bfseries 2022} (2022)
  083C01}.

\bibitem{Godfrey:1986wj}
S.~Godfrey and R.~Kokoski, ``{\em {The Properties of p Wave Mesons with One
  Heavy Quark}},'' \href{http://dx.doi.org/10.1103/PhysRevD.43.1679}{Phys. Rev.
  D {\bfseries 43} (1991) 1679--1687}.

\bibitem{Godfrey:2015dva}
S.~Godfrey and K.~Moats, ``{\em {Properties of Excited Charm and Charm-Strange
  Mesons}},'' \href{http://dx.doi.org/10.1103/PhysRevD.93.034035}{Phys. Rev. D
  {\bfseries 93} no.~3, (2016) 034035},
  \href{http://arxiv.org/abs/1510.08305}{[{\ttfamily 1510.08305}]}.

\bibitem{Godfrey:2016nwn}
S.~Godfrey, K.~Moats, and E.~S. Swanson, ``{\em {$B$ and $B_s$ Meson
  Spectroscopy}},'' \href{http://dx.doi.org/10.1103/PhysRevD.94.054025}{Phys.
  Rev. D {\bfseries 94} no.~5, (2016) 054025},
  \href{http://arxiv.org/abs/1607.02169}{[{\ttfamily 1607.02169}]}.

\bibitem{Wang:2022tuy}
G.-L. Wang, Q.~Li, T.~Wang, T.-F. Feng, X.-G. Wu, and C.-H. Chang, ``{\em {The
  solution to the \textquoteleft{}1/2 vs 3/2\textquoteright{} puzzle}},''
  \href{http://dx.doi.org/10.1140/epjc/s10052-022-10997-4}{Eur. Phys. J. C
  {\bfseries 82} no.~11, (2022) 1027},
  \href{http://arxiv.org/abs/2205.15470}{[{\ttfamily 2205.15470}]}.

\bibitem{Li:2018eqc}
Q.~Li, T.~Wang, Y.~Jiang, G.-L. Wang, and C.-H. Chang, ``{\em {Mixing angle and
  decay constants of $J^P=1^+$ heavy-light mesons}},''
  \href{http://dx.doi.org/10.1103/PhysRevD.100.076020}{Phys. Rev. D {\bfseries
  100} no.~7, (2019) 076020},
  \href{http://arxiv.org/abs/1802.06351}{[{\ttfamily 1802.06351}]}.

\bibitem{Sun:2014wea}
Y.~Sun, Q.-T. Song, D.-Y. Chen, X.~Liu, and S.-L. Zhu, ``{\em {Higher bottom
  and bottom-strange mesons}},''
  \href{http://dx.doi.org/10.1103/PhysRevD.89.054026}{Phys. Rev. D {\bfseries
  89} no.~5, (2014) 054026}, \href{http://arxiv.org/abs/1401.1595}{[{\ttfamily
  1401.1595}]}.

\bibitem{Lu:2016bbk}
Q.-F. L\"u, T.-T. Pan, Y.-Y. Wang, E.~Wang, and D.-M. Li, ``{\em {Excited
  bottom and bottom-strange mesons in the quark model}},''
  \href{http://dx.doi.org/10.1103/PhysRevD.94.074012}{Phys. Rev. D {\bfseries
  94} no.~7, (2016) 074012}, \href{http://arxiv.org/abs/1607.02812}{[{\ttfamily
  1607.02812}]}.

\bibitem{Gupta:1994mw}
S.~N. Gupta and J.~M. Johnson, ``{\em {Quantum chromodynamic potential model
  for light heavy quarkonia and the heavy quark effective theory}},''
  \href{http://dx.doi.org/10.1103/PhysRevD.51.168}{Phys. Rev. D {\bfseries 51}
  (1995) 168--175}, \href{http://arxiv.org/abs/hep-ph/9409432}{[{\ttfamily
  hep-ph/9409432}]}.

\bibitem{Vijande:2004he}
J.~Vijande, F.~Fernandez, and A.~Valcarce, ``{\em {Constituent quark model
  study of the meson spectra}},''
  \href{http://dx.doi.org/10.1088/0954-3899/31/5/017}{J. Phys. G {\bfseries 31}
  (2005) 481}, \href{http://arxiv.org/abs/hep-ph/0411299}{[{\ttfamily
  hep-ph/0411299}]}.

\bibitem{Yamada:2005nu}
Y.~Yamada, A.~Suzuki, M.~Kazuyama, and M.~Kimura, ``{\em {P-wave
  charmed-strange mesons}},''
  \href{http://dx.doi.org/10.1103/PhysRevC.72.065202}{Phys. Rev. C {\bfseries
  72} (2005) 065202}, \href{http://arxiv.org/abs/hep-ph/0601211}{[{\ttfamily
  hep-ph/0601211}]}.

\bibitem{Zhong:2008kd}
X.-h. Zhong and Q.~Zhao, ``{\em {Strong decays of heavy-light mesons in a
  chiral quark model}},''
  \href{http://dx.doi.org/10.1103/PhysRevD.78.014029}{Phys. Rev. D {\bfseries
  78} (2008) 014029}, \href{http://arxiv.org/abs/0803.2102}{[{\ttfamily
  0803.2102}]}.

\bibitem{Abreu:2019adi}
L.~M. Abreu, A.~G. Favero, F.~J. Llanes-Estrada, and A.~G. S\'anchez, ``{\em
  {Mixing and $m_q$ dependence of axial vector mesons in the Coulomb gauge QCD
  model}},'' \href{http://dx.doi.org/10.1103/PhysRevD.100.116012}{Phys. Rev. D
  {\bfseries 100} no.~11, (2019) 116012},
  \href{http://arxiv.org/abs/1908.11154}{[{\ttfamily 1908.11154}]}.

\bibitem{Ferretti:2015rsa}
J.~Ferretti and E.~Santopinto, ``{\em {Open-flavor strong decays of open-charm
  and open-bottom mesons in the $^3P_0$ model}},''
  \href{http://dx.doi.org/10.1103/PhysRevD.97.114020}{Phys. Rev. D {\bfseries
  97} no.~11, (2018) 114020},
  \href{http://arxiv.org/abs/1506.04415}{[{\ttfamily 1506.04415}]}.

\bibitem{Close:2005se}
F.~E. Close and E.~S. Swanson, ``{\em {Dynamics and decay of heavy-light
  hadrons}},'' \href{http://dx.doi.org/10.1103/PhysRevD.72.094004}{Phys. Rev. D
  {\bfseries 72} (2005) 094004},
  \href{http://arxiv.org/abs/hep-ph/0505206}{[{\ttfamily hep-ph/0505206}]}.

\bibitem{Chen:2016spr}
H.-X. Chen, W.~Chen, X.~Liu, Y.-R. Liu, and S.-L. Zhu, ``{\em {A review of the
  open charm and open bottom systems}},''
  \href{http://dx.doi.org/10.1088/1361-6633/aa6420}{Rept. Prog. Phys.
  {\bfseries 80} no.~7, (2017) 076201},
  \href{http://arxiv.org/abs/1609.08928}{[{\ttfamily 1609.08928}]}.

\bibitem{Aliev:2010ra}
T.~M. Aliev, A.~Ozpineci, and V.~Zamiralov, ``{\em {Mixing Angle of Hadrons in
  QCD: A New View}},''
  \href{http://dx.doi.org/10.1103/PhysRevD.83.016008}{Phys. Rev. D {\bfseries
  83} (2011) 016008}, \href{http://arxiv.org/abs/1007.0814}{[{\ttfamily
  1007.0814}]}.

\bibitem{Barnes:2002mu}
T.~Barnes, N.~Black, and P.~R. Page, ``{\em {Strong decays of strange
  quarkonia}},'' \href{http://dx.doi.org/10.1103/PhysRevD.68.054014}{Phys. Rev.
  D {\bfseries 68} (2003) 054014},
  \href{http://arxiv.org/abs/nucl-th/0208072}{[{\ttfamily nucl-th/0208072}]}.

\bibitem{Huang:2012ti}
Z.-W. Huang and J.~Liu, ``{\em {Analytic calculation of doubly heavy hadron
  spectral density in coordinate space}},''
  \href{http://arxiv.org/abs/1205.3026}{[{\ttfamily 1205.3026}]}.

\bibitem{Ioffe:2005ym}
B.~L. Ioffe, ``{\em {QCD at low energies}},''
  \href{http://dx.doi.org/10.1016/j.ppnp.2005.05.001}{Prog. Part. Nucl. Phys.
  {\bfseries 56} (2006) 232--277},
  \href{http://arxiv.org/abs/hep-ph/0502148}{[{\ttfamily hep-ph/0502148}]}.

\bibitem{Ioffe:2002ee}
B.~L. Ioffe, ``{\em {Condensates in quantum chromodynamics}},''
  \href{http://dx.doi.org/10.1134/1.1540654}{Phys. Atom. Nucl. {\bfseries 66}
  (2003) 30--43}, \href{http://arxiv.org/abs/hep-ph/0207191}{[{\ttfamily
  hep-ph/0207191}]}.

\bibitem{Narison:2022paf}
S.~Narison, ``{\em {QCD spectral sum rules 2022}},''
  \href{http://dx.doi.org/10.1016/j.nuclphysbps.2023.01.021}{Nucl. Part. Phys.
  Proc. {\bfseries 324-329} (2023) 94--106},
  \href{http://arxiv.org/abs/2211.14536}{[{\ttfamily 2211.14536}]}.

\end{thebibliography}\endgroup
